\documentclass[12pt,a4j]{article}
\usepackage[dvips]{graphicx}
\usepackage{enumerate}
\usepackage{amsmath}
\usepackage{amsfonts}

\begin{document}
\baselineskip=7mm
\centerline{\bf The bright $N$-soliton solution of a multi-component}\par
\centerline{\bf modified nonlinear Schr\"odinger equation  }\par
\bigskip
\centerline{Yoshimasa Matsuno\footnote{{\it E-mail address}: matsuno@yamaguchi-u.ac.jp}}\par

\centerline{\it Division of Applied Mathematical Science,}\par
\centerline{\it Graduate School of Science and Engineering} \par
\centerline{\it Yamaguchi University, Ube, Yamaguchi 755-8611, Japan} \par
\bigskip
\bigskip
\leftline{\bf Abstract}\par
\noindent A direct method is developed for constructing the bright  $N$-soliton solution of a multi-component modified nonlinear Schr\"odinger equation.
Specifically, the two different expressions of the solution are obtained both of which are expressed as a rational function of determinants.
A simple relation is found between them by employing the properties of the Cauchy matrix.
The proof of the solution reduces to the bilinear equations among the bordered determinants in which Jacobi's identity and related formulas
play a central role. 
Last, the bright $N$-soliton solution is presented for a (2+1)-dimensional nonlocal model equation  arising from the multi-component system as the number
of dependent variables tends to infinity.
  \par

\newpage
\leftline{\bf  1. Introduction} \par
\bigskip
\noindent The study of multi-component system of nonlinear partial differential equations (PDEs) is current interest in the theory of nonlinear waves.
Of particular concern are the  multi-component generalizations of the nonlinear Schr\"odinger (NLS) equation because of
their wide applicability in real physical systems such as nonlinear optics, nonlinear water waves and plasma physics [1-3].
The integrable two-component NLS system has been introduced for the first time by Manakov to describe the propagation of polarized electric field 
in an optical fiber and explored in detail by means of the
inverse scattering transform (IST) method [4]. After this remarkable work, various variants of integrable multi-component NLS systems have been proposed and analyzed
by the exact method of solution such as the IST and
Hirota's direct method.
In this paper, we consider the following multi-component system of  nonlinear PDEs which is a hybrid of the
coupled NLS equation and coupled derivative NLS equation
$${\rm i}\,q_{j,t}+q_{j,xx}+\mu\left(\sum_{k=1}^n|q_k|^2\right)q_j+{\rm i}\gamma\left[\left(\sum_{k=1}^n|q_k|^2\right)q_j \right]_x=0,\quad (j=1, 2, ..., n), \eqno(1.1)$$
where $q_j=q_j(x,t) \ (j=1, 2, ..., n)$ are  complex-valued functions of $x$ and $t$, $\mu$ and $\gamma$ are real constants, $n$ is an arbitrary positive integer
and subscripts $x$ and $t$ appended to $q_j$ denote  partial differentiations. 
The integrability of the above system has been established by constructing the Lax pair and an
infinite number of conservation laws [5]. For the two special cases $\gamma=0$ and $\mu=0$ reduced from the system (1.1) which correspond to  the multi-component NLS and multi-component derivative NLS
equations, respectively, their integrability has already been verified in [6, 7].
In the context of plasma physics, the two-component system with $\mu=0$ is a model equation for the propagation of polarized Alfv\'en waves. The single bright soliton
solution to this system has been obtained by means of the IST [8].
The two-component system with $\mu\not=0$ and $\gamma\not=0$   has been derived as a model for describing
the propagation of ultra-short pulses in birefringent optical fibers, together with its soliton solutions [9].
 Quite recently, we obtained the general bright  multisoliton solution (i.e., the  $N$-soliton solution  with $N$ being an arbitrary
positive integer which vanishes at infinity) of the two-component system by using a direct method [10]. 
It is important to remark that the constant $\mu$ must be nonnegative to support smooth bright solitons. The negative case is worth studying as well which will be treated in a separate issue.
\par
The purpose of this paper is to extend the results obtained in [10] to the general $n$-component system.
Specifically, we present the bright $N$-soliton solution of the $n$-component system (1.1) in the form of   compact determinantal expressions. 
Although the construction of the solution can be done following the
similar procedure to that developed in [10] for the two-component system, we provide a novel proof of the solution using  the expansion formulas for the bordered determinant.
\par
This paper is organized as follows. In section 2, we first transform the system (1.1) to a gauge equivalent system and then recast it to a
system of bilinear equations by introducing appropriate dependent variable transformations. Subsequently, the bright $N$-soliton solution to the bilinear equations is
presented. It has a simple structure expressed in terms of certain determinants. 
In section 3, we introduce some notations associated with the bright $N$-soliton solution and then prove several key formulas for determinants.  
In section 4, we perform the proof of the bright $N$-soliton solution using an elementary theory of determinants 
in which Jacobi's identity and related formulas will  play a central role.
In section 5, we provide an alternative
expression of the bright $N$-soliton solution and compare it with the corresponding solution presented in section 2.
We then find a simple relation between two types of solutions by employing the properties of the Cauchy matrix. 
This result leads to an alternative proof of the solution in a very simple way.
 In section 6, we discuss  a (2+1)-dimensional nonlocal modified NLS equation arising from the continuum limit $n \rightarrow \infty$
of the system (1.1). Specifically, we demonstrate that its bright $N$-soliton solution can be generated simply from that of the $n$-component system. Section 7 is devoted to 
 concluding remarks where we will comment on existing literatures about the bright $N$-soliton solutions of the multi-component NLS equation (equation (1.1) with $\gamma=0$)  and show that 
 the solutions obtained  in this paper include these known 
 solutions as special cases. \par
\newpage
\leftline{\bf 2. Bilinearization  and bright $N$-soliton solution}\par
\bigskip
\leftline{\it 2.1. Bilinearization}\par
\noindent We first apply the gauge transformations 
$$q_j=u_j\,{\rm exp}\left[-{{\rm i\gamma}\over 2}\int_{-\infty}^x\sum_{k=1}^n|u_k|^2dx\right],\quad (j=1, 2, ..., n), \eqno(2.1)$$
to the system (1.1) subjected to the  the boundary conditions $q_j\rightarrow 0, u_j\rightarrow 0$\ $(j=1, 2, ..., n)$ as $|x|\rightarrow \infty$,
where $u_j=u_j(x,t)\  (j=1, 2, ..., n)$ are complex-valued functions of $x$ and $t$.
Then, we obtain  the system of nonlinear PDEs for $u_j$
$${\rm i}\,u_{j,t}+u_{j,xx}+\mu\left(\sum_{k=1}^n|u_k|^2\right)u_j+{\rm i}\gamma\left(\sum_{k=1}^nu_k^*u_{k,x}\right)u_j=0,\quad (j=1, 2, ..., n), \eqno(2.2)$$
where the asterisk appended to $u_k$ denotes complex conjugate. 
This notation will be used frequently hereafter.
The second step in our analysis is given by the following proposition: \par
\bigskip
\noindent{\bf Proposition 2.1.}\ {\it By means of the dependent variable transformations
$$u_j={g_j\over f},\quad (j=1, 2, ..., n), \eqno(2.3)$$
the system of nonlinear PDEs (2.2) can be decoupled into the following system of bilinear equations for $f$ and $g_j$ 
$$({\rm i}D_t+D_x^2)g_j\cdot f=0, \quad (j=1, 2, ..., n), \eqno(2.4)$$
$$D_xf\cdot f^*={{\rm i}\gamma\over 2}\sum_{k=1}^n|g_k|^2, \eqno(2.5)$$
$$D_x^2f\cdot f^*=\mu\sum_{k=1}^n|g_k|^2+{{\rm i}\gamma\over 2}\sum_{k=1}^nD_xg_k\cdot g_k^*. \eqno(2.6)$$
Here, $f=f(x, t)$ and $g_j=g_j(x, t)\ (j=1, 2, ..., n)$ are complex-valued functions of $x$ and $t$ and
the bilinear operators $D_x$ and $D_t$ are defined by
$$D_x^mD_t^nf\cdot g=\left({\partial\over\partial x}-{\partial\over\partial x^\prime}\right)^m
\left({\partial\over\partial t}-{\partial\over\partial t^\prime}\right)^n
f(x, t)g(x^\prime,t^\prime)\Big|_{ x^\prime=x,\,t^\prime=t},  \eqno(2.7)$$
where $m$ and $n$ are nonnegative integers.} \par
\bigskip
\noindent {\bf Proof}.  Substituting (2.3) into (2.2) and rewriting the resultant equation in terms of the bilinear operators,
equations (2.2) can be rewritten as
$${1\over f^2}({\rm i}D_tg_j\cdot f+D_x^2g_j\cdot f)+{g_j\over f^3f^*}\left(-f^*D_x^2f\cdot f
+\mu f\sum_{k=1}^n|g_k|^2+{\rm i}\gamma\sum_{k=1}^ng_k^*D_xg_k\cdot f\right)=0, $$
$$ (j=1, 2, ..., n). \eqno(2.8)$$
Insert the identity
$$f^*D_x^2f\cdot f=fD_x^2f\cdot f^*-2f_xD_xf\cdot f^*+f(D_xf\cdot f^*)_x, \eqno(2.9)$$
into  the second term on the left-hand side of (2.8). Then, equations (2.8) become
$${1\over f^2}({\rm i}D_tg_j\cdot f+D_x^2g_j\cdot f)+{g_j\over f^3f^*}\Bigl[f\Bigl\{-D_x^2f\cdot f^*+\mu \sum_{k=1}^n|g_k|^2-(D_xf\cdot f^*)_x+{\rm i}\gamma\sum_{k=1}^ng_k^*g_{k,x}\Bigr\}$$
$$+f_x\Bigl\{2D_xf\cdot f^*-{\rm i}\gamma\sum_{k=1}^n|g_k|^2\Bigr\}\Bigr]=0, \quad (j=1, 2, ..., n). \eqno(2.10)$$
 As easily confirmed by a direct calculation,
 the left-hand side of (2.10) becomes zero by virtue of equations (2.4)-(2.6). \hspace{\fill}$\Box$
\par
\bigskip
It now follows from (2.3) and (2.5) that
$$-{{\rm i}\gamma\over 2}\sum_{k=1}^n|u_k|^2={\partial\over\partial x}\,{\rm ln}\,{f^*\over f}, \eqno(2.11)$$
which, substituted into (2.1), yields the solution of the system (1.1) in the form 
$$q_j={g_jf^*\over f^2},\qquad (j=1, 2, ..., n). \eqno(2.12)$$
Note that for the $n$-component NLS equation (the system (1.1) with $\gamma=0$), the solution (2.12) simplifies to $q_j=g_j/f$. Indeed,
if $\gamma=0$, then the bilinear equation (2.5) reduces to $D_xf\cdot f^*=0$. Thus, the ratio $f^*/f$ turns out to be an arbitrary function of $t$
which can be set to 1 under appropriate boundary condition. \par
\bigskip
\leftline{\it  2.2. Bright N-soliton solution} \par
\noindent We now state the main result in this paper:\par
\bigskip
\noindent {\bf Theorem 2.1.} {\it The bright $N$-soliton solution of the system of bilinear equations (2.4)-(2.6) is given by the
 determinants $f$ and $g_j\ (j=1, 2, ..., n)$ where
$$f=\begin{vmatrix} A & I\\ -I & B\end{vmatrix}, \quad g_j=\begin{vmatrix} A &I & {\bf z}^T \\ -I & B &{\bf 0}^T\\
                                                                            {\bf 0} & -{\bf a}_j^*
                                                                            & 0 \end{vmatrix},\quad (j=1, 2, ..., n). \eqno(2.13)$$
 Here, $A, B$ and $I$ are $N\times N$ matrices and ${\bf z}, {\bf a}_j$ and ${\bf 0}$ are $N$-component row vectors defined below and the symbol $T$ 
 denotes the transpose:
 $$A=(a_{jk})_{1\leq j,k\leq N}, \quad a_{jk}={1\over 2}\,{z_jz_k^*\over p_j+p_k^*}, \quad z_j={\rm exp}(p_jx+{\rm i}p_j^2t), \eqno(2.14a)$$
 $$B=(b_{jk})_{1\leq j,k\leq N}, \quad b_{jk}={(\mu+{\rm i}\gamma p_k)c_{jk}\over p_j^*+p_k}, \quad c_{jk}=\sum_{s=1}^n\alpha_{sj}\alpha^*_{sk},\eqno(2.14b)$$ 
$$I=(\delta_{jk})_{1\leq j,k\leq N}, : N\times N\ {\it unit\  matrix}, \eqno(2.14c)$$
$${\bf z}=(z_1, z_2, ..., z_N), \quad  {\bf a}_j=(\alpha_{j1}, \alpha_{j2}, ..., \alpha_{jN}), \quad
{\bf 0}=(0, 0, ..., 0). \eqno(2.14d)$$ } 
\par
The above bright $N$-soliton solution is characterized by  $(n+1)N$ complex parameters $p_j (j=1, 2, ..., n)$ and $\alpha_{sj} (s=1, 2, .., n;\ j=1, 2, ..., N).$
The former parameters determine the amplitude and velocity of the solitons whereas the latter ones determine the polarizations and the 
envelope phases of the solitons.  The conditions  
$p_j+p_k^*\not=0$ for all $j$ and $k$ and $p_j\not=p_k$ for $j\not=k$  may be imposed to assure the regularity of the solution. 
In the special case of $n=2$, (2.13) and (2.14) reproduce the bright $N$-soliton solution presented in [10].
The proof of Theorem 2.1 will be given in section 4. \par
To simplify the proof of theorem 2.1,  the following observation is useful: \par
\bigskip
\noindent {\bf Proposition 2.2.} {\it  If we introduce the gauge transformations
$$f=\tilde f, \quad g_j={\rm exp}\left[{\rm i}\left\{{\mu\over\gamma}\,\tilde x+\left({\mu\over\gamma}\right)^2\,\tilde t\right\}\right]\tilde g_j, \quad (j=1, 2, ..., n), \eqno(2.15a)$$
$$x=\tilde x+{2\mu\over\gamma}\,\tilde t, \quad t=\tilde t, \eqno(2.15b)$$
then the bilinear equations (2.4)-(2.6) recast to
$$({\rm i}D_{\tilde t}+D_{\tilde x}^2)\tilde g_j\cdot \tilde f=0, \quad (j=1, 2, ..., n), \eqno(2.16)$$
$$D_{\tilde x}\tilde f\cdot \tilde f^*={{\rm i}\gamma\over 2}\sum_{k=1}^n|\tilde g_k|^2, \eqno(2.17)$$
$$D_{\tilde x}^2\tilde f\cdot \tilde f^*={{\rm i}\gamma\over 2}\sum_{k=1}^nD_{\tilde x}\tilde g_k\cdot \tilde g_k^*, \eqno(2.18)$$
respectively. }\par
\bigskip
\noindent{\bf Proof.} The proof can be done by a straightforward calculation. \hspace{\fill}$\Box$ \par
\bigskip
Thus, the form of equations (2.4) and (2.5) is unchanged whereas equation (2.6) becomes a simplified equation with $\mu=0$. 
Consequently, the proof of the $N$-soliton solution may be performed for the corresponding solution with $\mu=0$. 
 Hence, in the analysis developed in the following sections, we put $\mu=0$ without loss of generality. \par
 \bigskip
\leftline{\bf 3. Notation and some basic formulas for determinants}\par
\bigskip
\noindent  In this section, we first introduce  the notation for matrices and then provide 
some basic formulas for determinants. \par
\bigskip
\noindent{\it 3.1. Notation}\par
\noindent We define the following matrices associated with the $N$-soliton solution (2.13) with (2.14):
$$D=\begin{pmatrix} A & I\\ -I & B\end{pmatrix},\eqno(3.1)$$
$$D({\bf a}^*;{\bf b})=\begin{pmatrix} A &I & {\bf 0}^T \\ -I & B &{\bf b}^T\\
                                                                            {\bf 0} & {\bf a}^*& 0 \end{pmatrix}, \quad
D({\bf a}^*;{\bf z})=\begin{pmatrix} A &I & {\bf z}^T \\ -I & B &{\bf 0}^T\\
                                                                            {\bf 0} & {\bf a}^*& 0 \end{pmatrix}, \quad
D({\bf z}^*;{\bf z})=\begin{pmatrix} A &I & {\bf z}^T \\ -I & B &{\bf 0}^T\\
                                                                            {\bf z}^* & {\bf 0}& 0 \end{pmatrix}. \eqno(3.2)$$
Note the position of the vectors ${\bf a}^*$, ${\bf b}, {\bf z}$ and ${\bf z}^*$ in the above expressions.
 The matrices which include more than two vectors
will be introduced as well. For example,
$$D({\bf a}^*,{\bf z}^*;{\bf b},{\bf z})=\begin{pmatrix} A &I& {\bf 0} & {\bf z}^T \\ -I & B &{\bf b}^T &{\bf 0}^T\\
                                                                            {\bf 0} & {\bf a}^*& 0 & 0 \\
              {\bf z}^* &{\bf 0} & 0& 0 \end{pmatrix}, \quad
D({\bf a}^*,{\bf z}^*;{\bf z},{\bf z}^\prime)=\begin{pmatrix} A &I& {\bf z}^T & {{\bf z}^\prime}^T \\ -I & B &{\bf 0}^T &{\bf 0}^T \\
                                                                            {\bf 0} & {\bf a}^*& 0 & 0 \\
              {\bf z}^* &{\bf 0} & 0& 0 \end{pmatrix}. \eqno(3.3)$$
 \par
\bigskip
\noindent{\it 3.2. Formulas for determinants}\par
\noindent Let $A=(a_{jk})_{1\leq j,k\leq M}$ be an $M\times M$  matrix with $M$ being an arbitrary 
positive integer,  $A_{jk}$ be the cofactor of the element $a_{jk}$ and ${\bf a}, {\bf b}, {\bf a}_j$ and ${\bf b}_j\ (j=1, 2, ..., n)$ be
$M$-component row vectors.
The following well-known formulas are used frequently  in our analysis [11]:
$${\partial\over\partial x}|A|=\sum_{j,k=1}^M{\partial a_{jk}\over\partial x}A_{jk}, \eqno(3.4)$$
$$\begin{vmatrix} A & {\bf a}^T\\ {\bf b} & z\end{vmatrix}=|A|z-\sum_{j,k=1}^MA_{jk}a_jb_k,  \eqno(3.5)$$
$$|A({\bf a}_1, {\bf a}_2; {\bf b}_1, {\bf b}_2)||A|= |A({\bf a}_1; {\bf b}_1)||A({\bf a}_2; {\bf b}_2)|-|A({\bf a}_1; {\bf b}_2)||A({\bf a}_2; {\bf b}_1)|. \eqno(3.6)$$
The formula (3.4) is the differentiation rule of the determinant and (3.5) is the expansion formula for a bordered determinant
with respect to the last row and last column.
The formula (3.6) is Jacobi's identity.   \par
The following two formulas may not be well-known but are very important in our analysis.
 In particular, formula (3.7) gives rise to the expansion formulas for the bordered determinant (see (3.9) and (3.10) below):
$$|A({\bf a}_1,..., {\bf a}_n; {\bf b}_1,..., {\bf b}_n)||A|^{n-1}=\begin{vmatrix} |A({\bf a}_1;{\bf b}_1)| &\cdots & |A({\bf a}_1;{\bf b}_n)| 
                                                                                    \\ \vdots & \ddots & \vdots                                                                                                                                                                                                                                                       
                                                                                    \\ |A({\bf a}_n;{\bf b}_1)| &\cdots & |A({\bf a}_n;{\bf b}_n)|
                                                                                     \end{vmatrix},\quad (n \geq 2), \eqno(3.7)$$
\begin{align} 
\left|A+\epsilon \sum_{s=1}^n{\bf b}_s^T{\bf a}_s\right| 
&= |A|+\sum_{m=1}^{n^\prime}(-\epsilon)^m\sum_{1\leq s_1<...<s_m\leq n}
|A({\bf a}_{s_1},...,{\bf a}_{s_m} ;{\bf b}_{s_1},...,{\bf b}_{s_m})| \notag \\
&=|A|+\sum_{m=1}^{n^\prime}{(-\epsilon)^m\over m!}\sum_{s_1,...,s_m=1}^n
|A({\bf a}_{s_1},...,{\bf a}_{s_m} ;{\bf b}_{s_1},...,{\bf b}_{s_m})|. \tag{3.8}
\end{align}
 Here, $\epsilon$ is an arbitrary parameter, the notation ${\bf b}_s^T{\bf a}_s$ on the left-hand side of (3.8) represents an $M\times M$ matrix whose $(j,k)$ element is given by $\beta_{sj}\alpha_{sk}$
 and $n^\prime={\rm min}(n, M)$.  The formula (3.7) is a variant of the Sylvester theorem  in
 the theory of determinants. 
 \par
 \bigskip
 \noindent{\bf Proof of (3.7).}   The proof proceeds by a mathematical induction.
 For $n=2$, (3.7) reduces to Jacobi's identity (3.6).  Assume that  formula (3.7) is true for $n-1\ (n\geq 3)$.  Let $L$ be the left-hand side of (3.7).
 Recall that the determinant changes its sign if any two rows (or columns) are interchanged. Applying this rule repeatedly to  $L$ when ${\bf a}_n$ and ${\bf b}_n$
 are shifted in front of ${\bf a}_1$ and ${\bf b}_1$, respectively, 
 $$L=|\hat A({\bf a}_1,..., {\bf a}_{n-1}; {\bf b}_1,..., {\bf b}_{n-1})||A|^{n-1},$$
 where $\hat A=A({\bf a}_n;{\bf b}_n)$ is an $(M+1)\times (M+1)$ matrix which is assumed to be nonsingular, i.e., $|\hat A|\not=0$.
  In view of the inductive hypothesis, $L$ can be written as
 $$L={|A|^{n-1}\over |\hat A|^{n-2}}\begin{vmatrix} |\hat A({\bf a}_1;{\bf b}_1)| &\cdots & |\hat A({\bf a}_1;{\bf b}_{n-1})| 
                                                                                    \\ \vdots & \ddots & \vdots                                                                                                                                                                                                                                                       
                                                                                    \\ |\hat A({\bf a}_{n-1};{\bf b}_1)| &\cdots & |\hat A({\bf a}_{n-1};{\bf b}_{n-1})|
                                                                                     \end{vmatrix}.$$
It follows from Jacobi's identity (3.6) that
$$|\hat A({\bf a}_j;{\bf b}_k)||A|=|A({\bf a}_n, {\bf a}_j; {\bf b}_n, {\bf b}_k)||A|=|A({\bf a}_n;{\bf b}_n)||A({\bf a}_j;{\bf b}_k)|-|A({\bf a}_j;{\bf b}_n)||A({\bf a}_n;{\bf b}_k)|.$$
 which, substituted into $L$, recasts $L$ into the form
 $$L={1\over |A({\bf a}_n;{\bf b}_n)|^{n-2}}\left|\Bigl(|A({\bf a}_n;{\bf b}_n)||A({\bf a}_j;{\bf b}_k)|-|A({\bf a}_j;{\bf b}_n)||A({\bf a}_n;{\bf b}_k)| \Bigr)_{1\leq j,k\leq n-1}\right|.$$
 Referring to the property of the bordered determinant, the above expression simplifies to
 $$L={1\over |A({\bf a}_n;{\bf b}_n)|^{n-2}}
 \begin{vmatrix}|A({\bf a}_n;{\bf b}_n)| |A({\bf a}_1;{\bf b}_1)| &\cdots & |A({\bf a}_n;{\bf b}_n)||A({\bf a}_1;{\bf b}_{n-1})| & |A({\bf a}_1;{\bf b}_n)|
 \\ \vdots & \ddots & \vdots & \vdots                                                                                                                                                                                                                                                       
 \\ |A({\bf a}_n;{\bf b}_n)||A({\bf a}_{n-1};{\bf b}_1)| &\cdots & |A({\bf a}_n;{\bf b}_n)||A({\bf a}_{n-1};{\bf b}_{n-1})| & |A({\bf a}_{n-1};{\bf b}_n)|
 \\ |A({\bf a}_n;{\bf b}_1)| &\cdots & |A({\bf a}_n;{\bf b}_{n-1})| & 1
 \end{vmatrix} $$
 Extract the factor $|A({\bf a}_n;{\bf b}_n)|$ from the $j$th row $(j=1, 2, ..., n-1)$ and then multiply  the last column  by the same factor.
 Then, the resultant expression is seen to be equal to the right-hand side of (3.7). \hspace{\fill}$\Box$ \par 
 \bigskip
 \noindent {\bf Proof of (3.8)}  
   For $n=1$, it follows by using the property of the bordered determinant that
\begin{align}
 \Big|A+\epsilon{\bf b}_1^T{\bf a}_1\Big|
 &=\begin{vmatrix} a_{11}+\epsilon\beta_{11}\alpha_{11} &\cdots & a_{1M}+\epsilon\beta_{11}\alpha_{1M} 
                                                                                    \\ \vdots & \ddots & \vdots                                                                                                                                                                                                                                                       
                                                                                    \\ a_{M1}+\epsilon\beta_{1M}\alpha_{11} &\cdots & a_{MM}+\epsilon\beta_{1M}\alpha_{1M}
                                                                                     \end{vmatrix} \notag \\
& =\epsilon\begin{vmatrix} a_{11} &\cdots & a_{1M} & \beta_{11}
                                                                                    \\ \vdots & \ddots & \vdots  &\vdots                                                                                                                                                                                                                                                      
                                                                                    \\ a_{M1} &\cdots & a_{MM} &\beta_{1M}
                                                                                    \\ -\alpha_{11} & \cdots & -\alpha_{1M} & {\epsilon}^{-1}
                                                                                     \end{vmatrix} 
 =\epsilon\begin{vmatrix}  A& {\bf b}_1^T \\
                   -{\bf a}_1 & {\epsilon}^{-1} \end{vmatrix}. \notag
 \end{align}
 Repeated use of the above modification yields
 $$\left|A+\epsilon \sum_{s=1}^n{\bf b}_s^T{\bf a}_s\right|=\epsilon^n\begin{vmatrix} A & {\bf b}_1^T &\cdots &{\bf b}_n ^T\\
                                                  -{\bf a}_1 &{\epsilon}^{-1}&\cdots& 0 \\
                                                  \vdots &  {\bf 0}^T & \ddots &{\bf 0}^T\\
                                                  -{\bf a}_n & 0& {\bf 0}& {\epsilon}^{-1} \end{vmatrix}. $$
 Expanding the determinant in powers of $\epsilon^{-1}$,  it is found that 
  $$\left|A+\epsilon \sum_{s=1}^n{\bf b}_s^T{\bf a}_s\right|=|A| +\epsilon^n\sum_{m=1}^n(-1)^m\sum_{1\leq s_1<...<s_m\leq n}
|A({\bf a}_{s_1},...,{\bf a}_{s_m} ;{\bf b}_{s_1},...,{\bf b}_{s_m})|\epsilon^{-(n-m)}. $$
The above expression coincides with (3.8) for $n\leq M$ since $n^\prime={\rm min}(n, M)=n$. For $M+1\leq n$, on the other hand,
 the determinant $|A({\bf a}_{s_1},...,{\bf a}_{s_m} ;{\bf b}_{s_1},...,{\bf b}_{s_m})|$ becomes zero identically for $M+1\leq m\leq n$, as 
confirmed by the Laplace expansion of the determinant  with respect to the last $m$ rows, for example. 
This implies that the summation with respect to $m$ is truncated at $m=M$ which is in accordance with (3.8) since $n^\prime={\rm min}(n, M)=M$.
 The second line of (3.8) follows from the facts that any permutation of the indices $\{s_1, s_2, ..., s_m\}$ does not alter the value of the
 determinant $|A({\bf a}_{s_1},...,{\bf a}_{s_m} ;{\bf b}_{s_1},...,{\bf b}_{s_m})|$ and the total number of the permutation is $m!$,
 and if the determinant includes at least two same rows (or columns), then it becomes zero identically. \hspace{\fill}$\Box$  \par
 \bigskip
  Suppose that $|A|\not=0$. Expanding the determinant on the right-hand side of (3.7)
 with respect to the first column and using (3.7) with $n$ replaced by $n-1$, we then obtain an expansion formula
 $$|A({\bf a}_1,..., {\bf a}_n; {\bf b}_1,..., {\bf b}_n)|={1\over |A|}\sum_{j=1}^n(-1)^{j-1}|A({\bf a}_j,{\bf b}_1)||A({\bf a}_1,...,{\bf a}_{j-1},{\bf a}_{j+1},..., {\bf a}_n; {\bf b}_2,..., {\bf b}_n)|.
  \eqno(3.9)$$
Similarly, the expansion with respect to the first row gives
$$|A({\bf a}_1,..., {\bf a}_n; {\bf b}_1,..., {\bf b}_n)|={1\over |A|}\sum_{j=1}^n(-1)^{j-1}|A({\bf a}_1,{\bf b}_j)||A({\bf a}_2,..., {\bf a}_n; {\bf b}_1,...,{\bf b}_{j-1},{\bf b}_{j+1},..., {\bf b}_n)|. \eqno(3.10)$$
\noindent The above two formulas will be used effectively to prove the bilinear equations (2.5) and (2.6). \par
\bigskip
 \noindent{\bf 4. Proof of the bright $N$-soliton solution}\par
 \bigskip
\noindent In this section, we show that the bright $N$-soliton solution (2.13) with (2.14) satisfies the system of bilinear equations (2.4)-(2.6).  
The proof can be performed for $\mu=0$, as noted at the end of section 2.  We first prove some formulas associated with the 
determinants $f$ and $g_j\ (j=1, 2, ..., n)$ and then
proceed to the proof. \par
\bigskip
\noindent {\it 4.1. Formulas}\par
\noindent In terms of the notation introduced in section 3.1\ (see (3.1) and (3.2)), $f$ and $g_j$ are written in the form
$$f=|D|, \qquad   g_j=-|D({\bf a}_j^*;{\bf z})|, \quad (j=1, 2, ..., n). \eqno(4.1)$$
The differentiation rules of $f$ and $g_j$ with respect to $t$ and $x$ are given by the following formulas: \par
\bigskip
\noindent{\bf Lemma 4.1.} \par
$$f_t=-{{\rm i}\over 2}\left\{|D({\bf z}^*;{\bf z}_x)|-|D({\bf z}_x^*;{\bf z})|\right\}, \eqno(4.2)$$
$$f_x=-{1\over 2}|D({\bf z}^*;{\bf z})|, \eqno(4.3)$$
$$f_{xx}=-{1\over 2}\left\{|D({\bf z}^*;{\bf z}_x)|+|D({\bf z}_x^*;{\bf z})|\right\}, \eqno(4.4)$$
$$g_{j,t}=-|D({\bf a}_j^*;{\bf z}_t)|+{{\rm i}\over 2}|D({\bf a}_j^*,{\bf z}^*;{\bf z},{\bf z}_x)|, \eqno(4.5)$$
$$g_{j,x}=-|D({\bf a}_j^*;{\bf z}_x)|, \eqno(4.6)$$
$$g_{j,xx}=-|D({\bf a}_j^*;{\bf z}_{xx})|+{1\over 2}|D({\bf a}_j^*,{\bf z}^*;{\bf z}_x,{\bf z})|. \eqno(4.7)$$
\noindent{\it Here, ${\bf z}_t$, ${\bf z}_x$ and ${\bf z}_{xx}$ are $N$-component row vectors given by
$ {\bf z}_t=({\rm i}p_1^2z_1, {\rm i}p_2^2z_2,..., {\rm i}p_N^2z_N)$, ${\bf z}_x=(p_1z_1, p_2z_2,..., p_Nz_N)$ and
${\bf z}_{xx}=(p_1^2z_1 ,p_2^2z_2,..., p_N^2z_N)$, respectively}. \par
\bigskip
\noindent{\bf Proof.} We prove (4.2). Let $D=(d_{jk})_{1\leq j,k\leq 2N}$ be a $2N\times 2N$ matrix and $D_{jk}$ be the cofactor of the element $d_{jk}$. 
It follows by applying the formula (3.4)
to the determinant $f$ given by (2.13) that 
\begin{align}
f_t &={{\rm i}\over 2}\sum_{j,k=1}^ND_{jk}(p_j-p_k^*)z_jz_k^* \notag \\
    &={{\rm i}\over 2}\sum_{j,k=1}^ND_{jk}(z_{j,x}z_k^*-z_jz_{k,x}^*), \notag
\end{align}
where in passing to the second line, use has been made of the relations $p_jz_j=z_{j,x},\ p_k^*z_k^*=z_{k,x}^*$. Referring to the formula (3.5) with $z=0$ and 
taking into account the notation (3.2), the above expression
reduces to the right-hand side of (4.2). 
A key feature in the proof is that the factor $(p_j+p_k^*)^{-1}$ in the element $a_{jk}$ has been canceled after differentiation with
respect to $t$. Using formulas (3.4) and (3.5) as well as some basic properties of determinants,
 formulas (4.3)-(4.7) can be proved  in the same way.  \hspace{\fill}$\Box$ \par 
 \bigskip
 The complex conjugate expressions of $f, f_x$ and $g_j$ can be expressed as follows:\par
 \bigskip
 \noindent{\bf Lemma 4.2.} \par
$$f^*=|\bar D|,\qquad \bar D\equiv \begin{pmatrix} A & I\\ -I & B-{\rm i}\gamma C\end{pmatrix},\eqno(4.8)$$
$$f_x^*=-{1\over 2}|\bar D({\bf z}^*;{\bf z})|, \eqno(4.9)$$
$$g_j^*=|\bar D({\bf z}^*;{\bf a}_j)|. \eqno(4.10)$$
\bigskip
\noindent{\bf Proof.} We prove (4.8).  It follows from (2.14a) and (2.14b) that $A^*=A^T$ and $B^*=B^T-{\rm i}\gamma C^T$ where
$C$ is an $N\times N$ matrix with elements $c_{jk}$  defined by $(2.14b)$. These relations lead to the expression of $f^*$
$$f^*=\begin{vmatrix} A^* & I\\ -I & B^*\end{vmatrix}=\begin{vmatrix} A^T & I\\ -I & (B-{\rm i}\gamma C)^T\end{vmatrix}.$$
Since $|A^T|=|A|$ for any square matrix $A$, the above expresion reduces to to the right-hand side of (4.8) after multiplying
the $j$th row and $k$th column $(j, k =1, 2, ...,N)$ by a facter $-1$.
Differentiating (4.8) with respect to $x$ and applying  formula (3.5) with $z=0$ to the resulting expression,  formula (4.9) follows immediately.
The proof of formula (4.10) can be done in the same way. \hspace{\fill}$\Box$ \par
The following formulas will be used in the proof of (2.5) and (2.6):\par
\bigskip
\noindent{\bf Lemma 4.3.} \par
$$|\bar D|=|D|+{1\over 2}|D({\bf z}^*;\tilde{\bf z})|, \eqno(4.11)$$
$$|D({\bf b}_k^*;\tilde{\bf z})|=|\bar D({\bf a}_k^*;{\bf z})|, \eqno(4.12)$$
$$|\bar D({\bf a}_k^*;{\bf b}_k)=-|D({\bf b}_k^*;{\bf a}_k)|-{1\over 2}|D({\bf b}_k^*,{\bf z}^*;{\bf a}_k,\tilde{\bf z})|, \eqno(4.13)$$
$$|\bar D({\bf a}_k^*;{\bf z}_x)|=|D({\bf b}_k^*;{\bf z})+{1\over 2}|D({\bf b}_k^*,{\bf z}^*;{\bf z},\tilde{\bf z})|. \eqno(4.14)$$

$$|D({\bf z}^*;{\bf z})|=2{\rm i}\gamma\sum_{k=1}^n|D({\bf b}_k^*;{\bf a}_k)|, \eqno(4.15)$$
$$|\bar D({\bf z}^*;{\bf z})|=-2{\rm i}\gamma\sum_{k=1}^n|\bar D({\bf a}_k^*;{\bf b}_k)|, \eqno(4.16)$$
{\it where $\tilde{\bf z}$ and ${\bf b}_k$ are $N$-component row vectors given respectively 
by $\tilde{\bf z}=(z_1/p_1, z_2/p_2, ..., z_N/p_N)$ and ${\bf b}_k=(\alpha_{k1}p_1^*, \alpha_{k2}p_2^*, ..., \alpha_{kN}p_N^*)$.} 
 \par
\bigskip
\noindent{\bf Proof.} First, we prove (4.11).  A direct calculation using the elements of $B$ and $C$ given by (2.14b) reveals that
$$b_{jk}-{\rm i}\gamma c_{jk}=-{p_j^*\over p_k}\,b_{jk}.$$
The determinant $|\bar D|$ from (4.8) is now modified to the form
$$|\bar D|=\begin{vmatrix} A & I\\ -I & \left(-{p_j^*\over p_k}\,b_{jk}\right) \end{vmatrix}
= \begin{vmatrix} \left(-{p_k^*\over p_j}\,a_{jk}\right) & I\\ -I & B\end{vmatrix},  $$
where the last line of the above expression  follows immediately from the property of the determinant. 
The definition (2.14a) of $a_{jk}$ now gives 
$$-{p_k^*\over p_j}\,a_{jk}=a_{jk}-{z_jz_k^*\over 2p_j}.$$
In view of the property of the bordered determinant,  $|\bar D|$ is modified in the form
$$|\bar D|=\begin{vmatrix} A & I & \tilde{\bf z}^T\\ -I & B & {\bf 0}^T \\ {{\bf z}^*\over 2} & {\bf 0} & 1\end{vmatrix}, $$
which is seen to coincide with (4.11) by applying formula (3.5). The proof of (4.12)-(4.14) can be done in the same way. Hence, we omit
the proof. \par
Let us now proceed to the proof of (4.15).
To this end, it is to be noted that the determinant $f$  can be rewritten in the form
$$f=\begin{vmatrix} \tilde A & I\\ -I & \tilde B\end{vmatrix}, $$
where $\tilde A$ and $\tilde B$ are $N\times N$ matrices defined by
$$\tilde A=(\tilde a_{jk})_{1\leq j,k\leq N}, \quad \tilde a_{jk}={1\over 2}\,{1\over p_j+p_k^*},  $$
 $$\tilde B=(\tilde b_{jk})_{1\leq j,k\leq N}, \quad \tilde b_{jk}={{\rm i}\gamma c_{jk}p_k\over p_j^*+p_k}\,z_j^*z_k.$$ 
 Indeed, the above expression of $f$ is derived from $f$ from (4.1) by extracting the factors $z_j$ and $z_k^*$ from the $j$th row and $k$th column, respectively
 for $j, k=1, 2, ..., N$ and then multiplying the $(N+j)$th row and $(N+k)$th column by the factors $z_j^*$ and $z_k$, respectively for $j, k=1, 2, ..., N$.
Using the formulas (3.4) and  (3.5) gives an alternative expression of $f_x$
$$f_x=-{\rm i}\gamma\sum_{k=1}^n|D({\bf b}_k^*;{\bf a}_k)|.$$
The formula (4.15) follows by comparing this expression with (4.3). The formula (4.16) comes from the complex conjugate expression of (4.15). \hspace{\fill}$\Box$ \par
\bigskip
\leftline{\it 4.2. Proof of (2.4)}\par
\noindent The proof of (2.4) proceeds following the same procedure as that of the same equation for $n=2$ [10]. 
Let $P_1$ be
$$P_1=({\rm i}D_t+D_x^2)g_j\cdot f. \eqno(4.17)$$
Substituting (4.1)-(4.7) into (4.17), $P_1$ becomes
$$P_1=-|D({\bf a}_j^*,{\bf z}^*;{\bf z},{\bf z}_x)||D|+|D({\bf a}_j^*;{\bf z})||D({\bf z}^*;{\bf z}_x)|-|D({\bf a}_j^*;{\bf z}_x)||D({\bf z}^*;{\bf z})|$$
$$-\left\{{\rm i}|D({\bf a}_j^*;{\bf z}_t)|+|D({\bf a}_j^*;{\bf z}_{xx})|\right\}. \eqno(4.18)$$
Referring to Jacobi's identity (3.6) and the fundamental formula 
$\alpha|D({\bf a};{\bf b}_1)|+\beta|D({\bf a};{\bf b}_2)|=|D({\bf a};\alpha{\bf b}_1+\beta{\bf b}_2)|$ \ $(\alpha, \beta \in \mathbb{C})$, $P_1$ simplifies to
$P_1=-|D({\bf a}_j^*;{\rm i}{\bf z}_t+{\bf z}_{xx})|$.
Since ${\rm i}\,{\bf z}_t+{\bf z}_{xx}={\bf 0}$ by $(2.14a)$, the last column of the  determinant consists only of zero elements, implying that $P_1=0.$ 
\hspace{\fill}$\Box$ \par
\bigskip
\leftline{\it 4.3. Proof of (2.5)}\par
\noindent The equation to be proved is $P_2=0$, where
$$P_2=D_xf\cdot f^*-{{\rm i}\gamma\over 2}\sum_{k=1}^n|g_k|^2. \eqno(4.19)$$
Substituting (4.1), (4.3) and (4.8)-(4.10) into (4.19), $P_2$ becomes
$$P_2=-{1\over 2}|\bar D||D({\bf z}^*;{\bf z})|+{1\over 2}|D||\bar D({\bf z}^*;{\bf z})|+{{\rm i}\gamma\over 2}\sum_{k=1}^n|D({\bf a}_k^*;{\bf z})||
\bar D({\bf z}^*;{\bf a_k})|. \eqno(4.20)$$
Further simplication is possible with use of
 (4.11), (4.15) and (4.16) with (4.13), giving rise to
$$P_2={{\rm i}\gamma\over 2}\sum_{k=1}^n\Bigl(-|D({\bf b}_k^*;{\bf a}_k)||D({\bf z}^*;\tilde{\bf z})|+|D({\bf b}_k^*,{\bf z}^*;{\bf a}_k,\tilde{\bf z})||D|
  +|D({\bf a}_k^*;{\bf z})||\bar D({\bf z}^*;{\bf a_k})|\Bigr). \eqno(4.21)$$
Applying Jacobi's identity (3.6) to the middle term 
and  replacing $|D({\bf b}_k^*;\tilde{\bf z})|$ by the right-hand side of (4.12) in the resultant expression, $P_2$ reduces to
$$P_2={{\rm i}\gamma\over 2}\sum_{k=1}^n\Bigl(-|\bar D({\bf a}_k^*;{\bf z})||D({\bf z}^*;{\bf a}_k)|+|D({\bf a}_k^*;{\bf z})||\bar D({\bf z}^*;{\bf a_k})|\Bigr). \eqno(4.22)$$
It now follows from (3.8) that
$$|\bar D({\bf a}_k^*;{\bf z})|=|D({\bf a}_k^*;{\bf z})|
+\sum_{m=1}^{n^{\prime\prime}}{\left({{\rm i}\gamma}\right)^m\over m!}\sum_{k_1,...,k_m=1}^n
|D({\bf a}_k^*,{\bf a}_{k_1}^*,...,{\bf a}_{k_m}^* ;{\bf z},{\bf a}_{k_1},...,{\bf a}_{k_m})|, \eqno(4.23a)$$
$$|\bar D({\bf z}^*;{\bf a}_k)|=|D({\bf z}^*;{\bf a}_k)|
+\sum_{m=1}^{n^{\prime\prime}}{\left({{\rm i}\gamma}\right)^m\over m!}\sum_{k_1,...,k_m=1}^n
|D({\bf z}^*,{\bf a}_{k_1}^*,...,{\bf a}_{k_m}^* ;{\bf a}_k,{\bf a}_{k_1},...,{\bf a}_{k_m})|, \eqno(4.23b)$$
where $n^{\prime\prime}={\rm min}(n-1,N-1)$.
 Referring to the expansion formulas (3.9) and (3.10), one has
 $$ |D({\bf a}_k^*,{\bf a}_{k_1}^*,...,{\bf a}_{k_m}^* ;{\bf z},{\bf a}_{k_1},...,{\bf a}_{k_m})|=|D|^{-1}|D({\bf a}_k^*;{\bf z})| |D({\bf a}_{k_1}^*,...,{\bf a}_{k_m}^* ;{\bf a}_{k_1},...,{\bf a}_{k_m})|$$
 $$+|D|^{-1}\sum_{l=1}^m(-1)^l|D({\bf a}^*_{k_l};{\bf z})||D({\bf a}_k^*,{\bf a}_{k_1}^*,...,{\bf a}^*_{k_{l-1}},{\bf a}^*_{k_{l+1}},...,{\bf a}_{k_m}^* ;{\bf a}_{k_1},...,{\bf a}_{k_m})|, \eqno(4.24a)$$
$$|D({\bf z}^*,{\bf a}_{k_1}^*,...,{\bf a}_{k_m}^* ;{\bf a}_k,{\bf a}_{k_1},...,{\bf a}_{k_m})|=|D|^{-1}|D({\bf z}^*;{\bf a}_k)| |D({\bf a}_{k_1}^*,...,{\bf a}_{k_m}^* ;{\bf a}_{k_1},...,{\bf a}_{k_m})|$$
$$+|D|^{-1}\sum_{l=1}^m(-1)^l|D({\bf z}^*;{\bf a}_{k_l})| |D({\bf a}_{k_1}^*,...,{\bf a}_{k_m}^* ;{\bf a}_k,{\bf a}_{k_1},...,{\bf a}_{k_{l-1}},{\bf a}_{k_{l+1}},...,{\bf a}_{k_m})|. \eqno(4.24b)$$
By introducing (4.23) into (4.22) and then using (4.24), $P_2$  takes the form
$$P_2={{\rm i}\gamma\over 2|D|}\sum_{m=1}^{n^{\prime\prime}}{\left({{\rm i}\gamma}\right)^m\over m!}\sum_{l=1}^m(-1)^l \times$$
$$\times \sum_{k,k_1,...,k_m=1}^n\Bigl[-|D({\bf a}^*_{k_l};{\bf z})||D({\bf z}^*;{\bf a}_k)|
|D({\bf a}_k^*,{\bf a}_{k_1}^*,...,{\bf a}^*_{k_{l-1}},{\bf a}^*_{k_{l+1}},...,{\bf a}_{k_m}^* ;{\bf a}_{k_1},...,{\bf a}_{k_m})|  $$
$$+|D({\bf a}_k^*;{\bf z})||D({\bf z}^*;{\bf a}_{k_l})||D({\bf a}_{k_1}^*,...,{\bf a}_{k_m}^* ;{\bf a}_k,{\bf a}_{k_1},...,{\bf a}_{k_{l-1}},{\bf a}_{k_{l+1}},...,{\bf a}_{k_m})|\Bigr]. \eqno(4.25)$$
Interchange the indices $k$ and $k_l$ in the first term and then shift the row vector ${\bf a}_{k_l}^*$  in front of ${\bf a}_{k_{l+1}}$ and  the column vector ${\bf a}_k$ in front of ${\bf a}_{k_1}$, respectively.
This leads to the following relation
$$|D({\bf a}_k^*,{\bf a}_{k_1}^*,...,{\bf a}^*_{k_{l-1}},{\bf a}^*_{k_{l+1}},...,{\bf a}_{k_m}^* ;{\bf a}_{k_1},...,{\bf a}_{k_m})|$$
\begin{align}
& \rightarrow |D({\bf a}_{k_l}^*,{\bf a}_{k_1}^*,...,{\bf a}^*_{k_{l-1}},{\bf a}^*_{k_{l+1}},...,{\bf a}_{k_m}^* ;{\bf a}_{k_1},...,{\bf a}_{k_{l-1}},{\bf a}_k,{\bf a}_{k_{l+1}},...,{\bf a}_{k_m})| \notag \\
&=|D({\bf a}_{k_1}^*,...,{\bf a}_{k_m}^* ;{\bf a}_k,{\bf a}_{k_1},...,{\bf a}_{k_{l-1}},{\bf a}_{k_{l+1}},...,{\bf a}_{k_m})|.\notag
\end{align}
Note that the value of the determinant is not altered since the total signature caused by the above manipulation is $(-1)^{2(l-1)}=1$. Thus, the first term 
on the right-hand side of (4.25) coincides with the second term and 
cosequently, $P_2=0$. \hspace{\fill}$\Box$ \par
\bigskip
\leftline{\it 4.3. Proof of (2.6)}\par
\noindent Instead of proving (2.6) directly, we differentiate (2.5) by  $x$ and add the resultant expression
 to (2.6) and then  prove the equation $P_3=0$, where
$$P_3=f_{xx}f^*-f_xf_x^*-{{\rm i}\gamma\over 2}\sum_{k=1}^ng_{k,x}g_k^*. \eqno(4.26)$$
This reduces the total amount of calculations considerably and the proof becomes transparent.
It now follows from (4.1), (4.3), (4.4), (4.6) and (4.8)-(4.10) that
$$P_3=-{1\over 2}\left\{|D({\bf z}^*;{\bf z}_x)|+|D({\bf z}_x^*;{\bf z})|\right\}|\bar D|-{1\over 4}|D({\bf z}^*;{\bf z})||\bar D({\bf z}^*;{\bf z})|
+{{\rm i}\gamma\over 2}\sum_{k=1}^n|D({\bf a}_k^*;{\bf z}_x)||\bar D({\bf z}^*;{\bf a}_k)|. \eqno(4.27)$$
Differention of (4.15) with respect to $x$ gives
$$|D({\bf z}^*;{\bf z}_x)|+|D({\bf z}_x^*;{\bf z})|=-{\rm i}\gamma\sum_{k=1}^n|D({\bf b}_k^*,{\bf z}^*;{\bf a}_k,{\bf z})|. \eqno(4.28)$$
Inserting  (4.15) and (4.28)  into (4.27), $P_3$ can be put into the form
$$P_3={{\rm i}\gamma\over 2}\sum_{k=1}^n\Bigl\{|\bar D||D({\bf b}_k^*,{\bf z}^*;{\bf a}_k,{\bf z})|+|D({\bf z}^*;{\bf z})||\bar D({\bf a}_k^*;{\bf b}_k)|
+|D({\bf a}_k^*;{\bf z}_x)||\bar D({\bf z}^*;{\bf a}_k)|\Bigr\}. \eqno(4.29)$$
Note from (4.11), (4.13), (4.14) and Jacobi's identity (3.6) that
$$|\bar D||D({\bf b}_k^*,{\bf z}^*;{\bf a}_k,{\bf z})|+|D({\bf z}^*;{\bf z})||\bar D({\bf a}_k^*;{\bf b}_k)|$$
\begin{align}
&=-|D({\bf z}^*;{\bf a}_k)|\Bigl\{|D({\bf b}_k^*;{\bf z})+{1\over 2}|D({\bf b}_k^*,{\bf z}^*;{\bf z},\tilde{\bf z})|\Bigr\} \notag \\ 
&=-|D({\bf z}^*;{\bf a}_k)||\bar D({\bf a}_k^*;{\bf z}_x)|. \tag{4.30}
\end{align}
After substituting (4.30) into (4.29), $P_3$ becomes
$$P_3={{\rm i}\gamma\over 2}\sum_{k=1}^n\Bigl\{-|\bar D({\bf a}_k^*;{\bf z}_x)||D({\bf z}^*;{\bf a}_k)|+|D({\bf a}_k^*;{\bf z}_x)||\bar D({\bf z}^*;{\bf a}_k)|\Bigr\}. \eqno(4.31)$$
This expression reduces to (4.22) if one replaces ${\bf z}_x$ by ${\bf z}$. Hence, the proof of the relation $P_3=0$ completely
 parallels  that of $P_2=0$ with $P_2$ from (4.22). \hspace{\fill}$\Box$ \par
\bigskip
\newpage
\leftline{\bf 5. Alternative expression of the bright $N$-soliton solution}\par
\bigskip
\noindent Here, we persent an alternative expression of the bright $N$-soliton solution in terms of  the determinants with smaller sizes when compared with those given by (2.13).
Explicitly, we write it as a theorem:\par
\bigskip
\noindent {\bf Theorem 5.1.} {\it The determinants $f^\prime$ and $g_j^\prime\ (j=1, 2, ..., n)$  given below satisfy the system of bilinear equations (2.4)-(2.6): \par
$$f^\prime=|A^\prime+B^\prime|, \quad g_j^\prime=\begin{vmatrix} A^\prime +B^\prime &  {\bf y}^T \\   -{{\bf a}_j^\prime}^* & 0 \end{vmatrix},\qquad  (j=1, 2, ..., n), \eqno(5.1)$$
where  $A^\prime$ and  $ B^\prime$  are $N\times N$ matrices and ${\bf y}$ and ${\bf a}_j^\prime$ are $N$-component row vectors defined below: \par
 $$A^\prime=(a_{jk}^\prime)_{1\leq j,k\leq N}, \quad a_{jk}^\prime={1\over 2}\,{y_jy_k^*\over q_j+q_k^*}, \quad y_j={\rm exp}(q_jx+{\rm i}q_j^2t), \eqno(5.2a)$$
 $$B^\prime=(b_{jk}^\prime)_{1\leq j,k\leq N}, \quad b_{jk}^\prime={(\mu-{\rm i}\gamma q_k^*)c_{jk}^\prime\over q_j+q_k^*}, \quad c_{jk}^\prime=\sum_{s=1}^n\alpha_{sj}^\prime{\alpha^\prime_{sk}}^*,\eqno(5.2b)$$ 
 $${\bf y}=(y_1, y_2, ..., y_N),\qquad {\bf a}_j^\prime=(\alpha_{j1}^\prime, \alpha_{j2}^\prime, ..., \alpha_{jN}^\prime). \eqno(5.2c)$$
 Here, $q_j\ (j=1, 2, ..., N)$ and $\alpha_{sj}^\prime\ (s=1, 2, ..., n; j=1, 2, ..., N)$ are complex parameters characterizing the solution. } \par
\bigskip
\noindent{\bf Proof.} The proof of the solution can be performed in the same way as that of (2.13) with (2.14). Indeed, the proof of (2.4), (2.5) and (2.6) reduce respectively to the relations (4.18), (4.22)
and (4.31) in which the matrix $D$ may  be replaced simply by the matrix $A^\prime+B^\prime$. \hspace{\fill}$\Box$ \par
\bigskip
Let us  show that the determinants $f$ and $g_j$ from (2.13)  are  closely related to the determinants  $f^\prime$ and $g_j^\prime$ given by (5.1).  
The following lemma is useful for this purpose:\par
\bigskip
\noindent{\bf Lemma 5.1.} {\it The determinants $f$ and $g_j$ given by (2.13) can be rewritten in the form
$$f=|I+AB|,\qquad g_j=\begin{vmatrix}I+AB &  {\bf z}^T \\   -{\bf a}_j^* & 0 \end{vmatrix},\qquad  (j=1, 2, ..., n). \eqno(5.3)$$ }
\bigskip
\noindent{\bf Proof.} Multiplying $f$ from (2.13) by a factor $\begin{vmatrix} B & -I\\ I & O\end{vmatrix}$  and performing the operation
of matrix multiplication, the first expression of (5.3) follows immediately. Similarly, the second expression is obtained if one multiplies $g_j$ by a
factor $\begin{vmatrix} B &-I & {\bf 0}^T \\ I & O &{\bf 0}^T\\ {\bf 0} & {\bf 0} & 1\end{vmatrix}$. Indeed,
$$\begin{vmatrix} A & I\\ -I & B\end{vmatrix}\begin{vmatrix} B & -I\\ I & O\end{vmatrix}=\begin{vmatrix} I+AB & -A\\ O & I\end{vmatrix}=|I+AB|,$$
$$\begin{vmatrix} A &I & {\bf z}^T \\ -I & B &{\bf 0}^T\\{\bf 0} & -{\bf a}_j^* & 0 \end{vmatrix}\begin{vmatrix} B &-I & {\bf 0}^T \\ I & O &{\bf 0}^T\\ {\bf 0} & {\bf 0} & 1\end{vmatrix}
= \begin{vmatrix} I+AB &-A& {\bf z}^T \\ O & I &{\bf 0}^T\\ -{\bf a}_j^* & {\bf 0} & 0 \end{vmatrix}=\begin{vmatrix} I+AB &  {\bf z}^T \\   -{\bf a}_j^* & 0 \end{vmatrix}. $$ 
Since the value of each factor multiplied from the right is 1, (5.3) follows. \hspace{\fill}$\Box$ \par
\bigskip
We now establish the following theorem:\par
\bigskip
\noindent{\bf Theorem 5.2.} {\it Under the parameterization $q_j=-p_j^*\ (j=1, 2, ..., N)$ and $\alpha_{sj}^\prime=-\alpha_{sj}/(2c_j^*)
\ (s=1, 2, ..., n; j=1, 2, ..., N)$, 
the determinants $f, f^\prime, g_j$ and $g_j^\prime$ satisfy the relations
$$f=c|A|f^\prime, \eqno(5.4)$$
$$g_j=c|A|g_j^\prime, \quad (j=1, 2, ..., n), \eqno(5.5)$$
where 
$$c=(-1)^N \prod_{l=1}^N(4c_l^*c_l),\qquad c_l={\prod_{m=1}^N(p_l+p_m^*)\over \prod_{\substack{m=1\\(m\not=l)}}^N(p_l-p_m)}, \qquad (l= 1, 2, ..., N). \eqno(5.6)$$
The parameters $p_j\ (j=1, 2, ..., N)$ are assumed to satisfy the conditions $p_l+p_m^*\not=0$ for all $l$ and $m$ and $p_l\not=p_m$ for $l\not=m$. } \par
 \bigskip
\noindent{\bf Proof}. Let $\tilde A$ be a Cauchy matrix of the form $\tilde A=\left({1\over 2}{1\over p_j+p_k^*}\right)$. 
Then, $A=(z_j\delta_{jk})\tilde A (z_j^*\delta_{jk})$. Since $|\tilde A|\not=0$  by virtue of the well-known formula for $|\tilde A|$ and 
the conditions imposed on the parameters $p_j\ (j=1, 2, ..., N)$, the inverse of
$\tilde A$ exists, implying that $A^{-1}$ exists as well. Actually, it reads
$A^{-1}=(z_j^*\delta_{jk})^{-1}\tilde A^{-1} (z_j\delta_{jk})^{-1}$. Using the explicit expression of $\tilde A^{-1}$, i.e., $\left({2c_j^*c_k\over p_j^*+p_k}\right)$ [11], 
the inverse matrix $A^{-1}$ can be written in the form
$$A^{-1}=\left({2c_j^*c_k\over p_j^*+p_k}{1\over z_j^*z_k}\right). \eqno(5.7)$$
Applying the basic properties of determinants to $f$ and $g_j$ from (5.3) gives
$$f=|A||A^{-1}+B|,\eqno(5.8)$$
$$g_j=|A|\begin{vmatrix}A^{-1}+B &  A^{-1}{\bf z}^T \\   -{\bf a}_j^* & 0 \end{vmatrix},\qquad  (j=1, 2, ..., n). \eqno(5.9)$$
The $j$th element of the column vector $A^{-1}{\bf z}^T$ is
\begin{align} (A^{-1}{\bf z}^T)_j
&=\sum_{l=1}^N(A^{-1})_{jl}z_l \notag \\
&={2c_j^*\over z_j^*}\sum_{l=1}^N{1\over p_j^*+p_l}{\prod_{m=1}^N(p_l+p_m^*)\over \prod_{\substack{m=1\\(m\not=l)}}^N(p_l-p_m)}\notag \\
&={2c_j^*\over z_j^*}\sum_{l=1}^N{\prod_{\substack{m=1\\(m\not=j)}}^N(p_l+p_m^*)\over \prod_{\substack{m=1\\(m\not=l)}}^N(p_l-p_m)}.\tag{5.10}
\end{align}
By Euler's formula, the sum in the last line turns out to be 1 and hence $(A^{-1}{\bf z}^T)_j=2c_j^*/z_j^*$. Introducing this relation into $g_j$,
$$g_j=|A|\begin{vmatrix}A^{-1}+B &\hat{\bf z}^T\\   -{\bf a}_j^* & 0 \end{vmatrix},\qquad  (j=1, 2, ..., n), \eqno(5.11) $$
where $\hat{\bf z}=(2c_1^*/z_1^*, 2c_2^*/z_2^*, ..., 2c_N^*/z_N^*)$ is an $N$-component row vector. 
\par
The next step is to modify the determinants $f^\prime$ and $g_j^\prime$. By means of the parametrization $q_j=-p_j^*$, $y_j$ from (5.2a) is related to $z_j$ from (2.14a) by the relation $y_j={z_j^*}^{-1}$.
Similary, the relation $c_{jk}^\prime=c_{jk}/(4c_j^*c_k)$  follows from (2.14b) and (5.2b)  and the parameterization $\alpha_{sj}^\prime=-\alpha_{sj}/(2c_j^*)$.
Substitution of these relations into $f^\prime$  gives
\begin{align} f^\prime
&=\left|\left({1\over 2}{y_jy_k^*\over q_j+q_k^*}+{(\mu-{\rm i}\gamma q_k^*)c_{jk}^\prime\over q_j+q_k^*}\right)\right| \notag \\
&={(-1)^N\over \prod_{l=1}^N(4c_l^*c_l)}\left|\left({2c_j^*c_k\over p_j^*+p_k}{1\over z_j^*z_k}+{(\mu+{\rm i}\gamma p_k)c_{jk}\over p_j^*+p_k}\right)\right| \notag \\
&=c^{-1}|A^{-1}+B|, \tag{5.12}
\end{align}
where in passing to the second line,  the factor $1/(2c_j^*)$ has been extracted from the $j$th row $(j=1, 2, .., N)$ and the factor $-1/(2c_k)$ from the  $k$th column $(k=1, 2, .., N)$, respectively.
The similar procedure applied to $g_j^\prime$ leads to the expression
$$g_j^\prime=c^{-1}\begin{vmatrix}A^{-1}+B &\hat{\bf z}^T\\    -{\bf a}_j^* & 0 \end{vmatrix},\qquad  (j=1, 2, ..., n). \eqno(5.13) $$
The relation (5.4) follows from (5.8) and (5.12) whereas the relation (5.5) follows from (5.11) and (5.13). \hspace{\fill}$\Box$ \par
\bigskip
Thus, we have obtained the two different expressions for the bright $N$-soliton solution of the system of nonlinear PDEs (2.2). Explicitly, they read $u_j=g_j/f=g_j^\prime/f^\prime\ (j=1, 2, ..., n).$
\par
The following proposition provides an alternative proof of theorem 5.1:\par
\bigskip
\noindent {\bf Proposition 5.1.} {\it If $f$ and $g_j$ given respectively by (5.4) and (5.5) satisfy the system of bilinear equations (2.4)-(2.6), then $f^\prime$ and $g_j^\prime$ satisfy the 
same system of equations, and vice versa.} \par
\bigskip
\noindent {\bf Proof.} Substituting (5.4) and (5.5) into (2.4) and using the definition of the bilinear operators, 
$$c^2|A|^2({\rm i}D_tg_j^\prime\cdot f^\prime+D_x^2g_j^\prime\cdot f^\prime)+c^2(D_x^2|A|\cdot|A|)g_j^\prime f^\prime=0. \eqno(5.14)$$
 The Cauchy type determinant $|A|$ can be modified, after  extracting the factor $z_j$ from $j$th row $(j=1, 2, ..., N)$ and the factor $z_k^*$ from $k$th column $(k=1, 2, ..., N)$, respectively,  in the form
\begin{align} |A|&= \prod_{l=1}^N(z_lz_l^*)\left|\left({1\over 2}{1\over p_j+p_k^*}\right)\right| \notag \\
&={\rm exp}\left[\sum_{l=1}^N(p_l+p_l^*)x+{\rm i}\sum_{l=1}^N(p_l^2-{p_l^*}^2)t\right]\left|\left({1\over 2}{1\over p_j+p_k^*}\right)\right|. \tag{5.15}
\end{align}
Differentiation of $|A|$ with respect to $x$ gives
$$|A|_x=\sum_{l=1}^N(p_l+p_l^*)|A|, \qquad |A|_{xx}=\left\{\sum_{l=1}^N(p_l+p_l^*)\right\}^2|A|. \eqno(5.16)$$
It immediately follows from (5.16) that
$$D_x^2|A|\cdot |A|=2(|A||A|_{xx}-|A|_x^2)=0. \eqno(5.17)$$
It is seen from (5.14), (5.17) and the relation $c|A|\not=0$ that $f^\prime$ and $g_j^\prime$ satisfy the bilinear equation (2.4). 
The remaining part of the proposition can be proved in the same way if one uses (5.17) and the reality of $|A|$, i.e., $|A|^*=|A|$ 
which is a consequence of the Hermitian nature of the matrix $A$. 
The proof of the converse proposition  proceeds in the same way if one uses the relation $D_x^2|A|^{-1}\cdot |A|^{-1}=0$ in place of (5.17).
\hspace{\fill}$\Box$ \par
\bigskip
\leftline{\bf 6. A continuum model} \par
\bigskip
\noindent The $n$-component system (1.1) yields a continuum model when one takes a limit $n\rightarrow \infty$. 
It represents a (2+1)-dimensional nonlocal modified NLS equation of the form
$${\rm i}\,q_t+q_{xx}+\mu\left(\int_{-\infty}^\infty|q|^2dy\right)q+{\rm i}\gamma \left(\int_{-\infty}^\infty|q|^2dy\, q\right)_x=0, \quad q=q(x,y,t). \eqno(6.1)$$
Recall that when $\gamma=0$, this equation reduces to a (2+1)-dimensional nonlocal NLS equation proposed by Zakharov [12].  The exact method of solution
for equation (6.1) can be developed following the same procedure as that for the system of nonlinear PDEs (1.1). Hence, we summarize the main results. \par
First, application of the gauge transformation
$$q=u\,{\rm exp}\left[-{{\rm i\gamma}\over 2}\int_{-\infty}^x\int_{-\infty}^\infty|u(x,y,t)|^2dxdy\right],\qquad u=u(x,y,t), \eqno(6.2)$$
to the system (6.1) subjected to the  the boundary conditions $q\rightarrow 0, u\rightarrow 0$ $|x|\rightarrow \infty$ transforms (6.1) to a nonlocal nonlinear PDE for $u$
$${\rm i}\,u_{t}+u_{xx}+\mu\left(\int_{-\infty}^\infty|u|^2dy\right)u+{\rm i}\gamma\left(\int_{-\infty}^\infty u^*u_{x}dy\right)u=0. \eqno(6.3)$$
The proposition below is an analog of proposition 2.1:\par
\bigskip
\noindent{\bf Proposition 6.1} {\it By means of the dependent variable transformation
$$u={g\over f}, \eqno(6.4)$$
 equation (6.3) can be decoupled into the following system of bilinear equations for $f=f(x,t)$ and $g=g(x,y,t)$ 
$$({\rm i}D_t+D_x^2)g\cdot f=0, \eqno(6.5)$$
$$D_xf\cdot f^*={{\rm i}\gamma\over 2}\int_{-\infty}^\infty|g|^2dy, \eqno(6.6)$$
$$D_x^2f\cdot f^*=\mu\int_{-\infty}^\infty|g|^2dy+{{\rm i}\gamma\over 2}\int_{-\infty}^\infty D_xg\cdot g^*dy. \eqno(6.7)$$ } \par
\bigskip
{\noindent {\bf Proof.} The proof proceeds exactly as that of proposition 2.1. Formally, one may simply replace the sum $\sum_{k=1}^n$ by the integral $\int_{-\infty}^\infty dy$.  \hspace{\fill}$\Box$ \par
\bigskip
It follows from (6.2), (6.4) and (6.6) that
$$q={gf^*\over f^2}, \eqno(6.8)$$
which is just a continuum limit of (2.13).
\par
The following theorem can be derived from a continuum limit of the bright $N$-soliton solution given by theorem 2.1 and theorem 5.1: \par
\bigskip
\noindent {\bf Theorem 6.1.} {\it  The system of bilinear equations (6.5)-(6.7) admits the following two different expressions $f, g$ and $f^\prime, g^\prime$ for the bright $N$-soliton solution : 
$$f=\begin{vmatrix} A & I\\ -I & B\end{vmatrix}, \quad g=\begin{vmatrix} A &I & {\bf z}^T \\ -I & B &{\bf 0}^T\\
                                                                            {\bf 0} & -{\bf a}^*
                                                                            & 0 \end{vmatrix}, \eqno(6.9)$$
$$f^\prime=|A^\prime+B^\prime|, \quad g^\prime=\begin{vmatrix} A^\prime +B^\prime &  {\bf y}^T \\   -{{\bf a}^\prime}^* & 0 \end{vmatrix}. \eqno(6.10)$$
                                                                            
\noindent Here, $A$ and  $B$  are $N\times N$ matrices given respectively by (2.14a) and (2.14b) with $c_{jk}$ being replaced by $\int_{-\infty}^\infty\alpha_j(y)\alpha_k^*(y)dy$,
  $A^\prime$ and $B^\prime$  are $N\times N$ matrices given respectively by (5.2a) and (5.2b) with $c_{jk}^\prime$ being replaced by $\int_{-\infty}^\infty\alpha_j^\prime(y){\alpha_k^\prime}^*(y)dy$ 
 and
 ${\bf a}={\bf a}(y)=(\alpha_1, \alpha_2, ..., \alpha_N)$  and  ${\bf a}^\prime={\bf a}^\prime(y)=(\alpha_1^\prime, \alpha_2^\prime, ..., \alpha_N^\prime)$
 are $N$-component row vectors where  $\alpha_j$ and $\alpha_j^\prime\ (j=1, 2, ..., N)$ are continuous functions of $y$. }
  \par
 \bigskip
  \noindent {\bf Proof.} The proof can be done in the same way as that of theorem 2.1 and theorem 5.1. \hspace{\fill}$\Box$ \par
 \bigskip
 \noindent{\bf Theorem 6.2.} {\it Under the parameterization $q_j=-p_j^*$ and $\alpha_{j}^\prime=-\alpha_{j}/(2c_j^*)$ 
 \ $(j=1, 2, ..., N)$, 
the determinants $f, f^\prime, g$ and $g^\prime$ satisfy the relations
$$f=c|A|f^\prime, \eqno(6.11)$$
$$g=c|A|g^\prime, \eqno(6.12)$$
where $c$ is defined by (5.6) and
the parameters $p_j\ (j=1, 2, ..., N)$ are specified such that  $p_l+p_m^*\not=0$ for all $l$ and $m$ and $p_l\not=p_m$ for $l\not=m$. } \par
 \bigskip
 \noindent {\bf Proof.} The proof parallels  theorem 5.2. \hspace{\fill}$\Box$ \par
 \bigskip
 \noindent {\bf Proposition 6.2.} {\it If $f$ and $g$ given  by (6.9)  satisfy the system of bilinear equations (6.5)-(6.7), then $f^\prime$ and $g^\prime$ given by (6.11) and (6.12)
  satisfy the 
same system of equations, and vice versa.} \par
\bigskip
\noindent {\bf Proof.} The proof is completely parallel to that of proposition 5.1. \hspace{\fill}$\Box$ \par
\bigskip
 \noindent{\bf 7. Concluding remarks}\par
 \bigskip
 \noindent In this paper, we have obtained  the two different expressions for the bright $N$-soliton solution of an $n$-component modified NLS equation
 and found a simple relationship between them. 
 We also have presented the  bright $N$-soliton solution of a continuum model arising from the system 
 as the number $n$ of the dependent variables tends to infinity. These solutions include, as special cases, 
  existing solutions for a  multi-component
 NLS equation. Actually, when $\gamma=0$, the system of nonlinear PDEs (1.1) reduces to an   $n$-component  NLS equation. It admits the
 bright $N$-soliton solution of the form $q_j=g_j/f=g_j^\prime/f^\prime\ (j=1, 2, .., n).$ 
 This fact follows from (2.12) and the relations $f^*=f$ and ${f^\prime}^*=f^\prime$. See (2.13) and (5.1). 
 The solution (2.13) with $\gamma=0$ has been obtained in [13] using a direct method 
 whereas the solution  (5.3) with $\gamma=0$ has been constructed by means of the IST [14]. An alternative expression (5.1) of the solution with $\gamma=0$ has been
 derived by employing a method of algebraic geometry [15]. 
 For a continuum model (6.1) with $\gamma=0$, the solution takes the form $q=g/f$.
 The bright $N$-soliton solution  (6.9) with $\gamma=0$
 has been found in [16] by a direct method. \par
 In a future work, we will investigate the various features of the multi-component bright solitons. In particular, we will be concerned with
 the effect of the parameters $\mu$ and $\gamma$ which characterize the different types of nonlinearities on the interaction process of solitons. \par
 \bigskip
                                                                    
\leftline{\bf Acknowledgement}\par
\bigskip
This work was partially supported by the Grant-in-Aid for Scientific Research (C) No. 22540228 from Japan Society for the Promotion of Science. \par

\newpage

\leftline{\bf References}\par
\begin{enumerate}[{[1]}]
\item Hasegawa A and Kodama Y 1995 {\it Solitons in Optical Communications } (New York: Oxford)
\item Kivshar Y S and Agrawal G P 2003 {\it Optical Solitons From Fibers to Photonic Crystals} (New York: Academic)
\item Maimistov A I 2010 { Solitons in nonlinear optics} {\it Quantum Electron.} {\bf 40} 756-781
\item Manakov S V 1974 { On the theory of two-dimensional stationary self-focusing of electromagnetic waves} {\it Sov. Phys. - JETP} {\bf 38} 248-253
\item Hisakado M and Wadati M 1995  Integrable multi-component hybrid nonlinear Schr\"odinger equations {\it J. Phys. Soc. Japan} {\bf 64} 408-413
\item Makhan'kov V G and Pashaev O K 1982 { Nonlinear  Schr\"odinger equation with noncompact isogroup} {\it Theor. Math. Phys.} {\bf 53} 979-987
\item Fordy A P 1984 { Derivative nonlinear Schr\"odinger equations and Hermitian symmetric spaces} {\it J. Phys. A: Math. Gen.} {\bf 17} 1235-1245
\item Morris H C and Dodd R K 1979 {The two component derivative nonlinear Schr\"odinger equation} {\it Phys. Scr.} {\bf 20} 505-508
\item Hisakado M, Iizuka T and Wadati M 1994 { Coupled hybrid nonlinear Schr\"odinger equation and optical solitons} {\it J. Phys. Soc. Japan} {\bf 63} 2887-2894
\item Matsuno Y 2011 { The $N$-soliton solution of a two-component modified nonlinear Schr\"odinger equation} {\it Phys. Lett.} {\bf A 375} 2090-2094
\item Vein R and Dale P 1999 {\it Determinants and Their Applications in Mathematical Physics} (New York: Springer)
\item Zakharov V E 1980 {The inverse scattering method} {\it Solitons (Topics in Current Physics)} ed R K Bullough and P J Caudrey (Berlin: Springer) pp 243-285
\item Ablowitz M J, Ohta Y and Trubatch A D 1999 { On discretizations of a vector nonlinear Schr\"odinger equation} {\it Phys. Lett.} {\bf A 253} 287-304
\item Tsuchida T 2004 {$N$-soliton collision in the Manakov model} {\it Prog. Theor. Phys.} {\bf 111} 151-182
\item Dubrovin B A, Malanyuk T M, Krichever I M and Makhan'kov V G 1988 { Exact solutions of the time-dependent Schr\"odinger equation
                                                                  with self-consistent potentials} {\it Sov. J. Part. Nucl.} {\bf 19} 252-269
\item Maruno K and Ohta Y 2008 { Localized solutions of a (2+1)-dimensional nonlocal nonlinear Schr\"odinger equation} {\it Phys. Lett.} {\bf A 372} 4446-4450

\end{enumerate}

\end{document}